%% file: a222+3.tex
\documentclass[11pt]{article}

\usepackage{graphicx}
\usepackage{natbib}
\bibpunct{(}{)}{;}{a}{}{,}


\def\arcmin{\hbox{$^\prime$}}

\def\farcs{\hbox{$.\!\!^{\prime\prime}$}}

\def\sun{\hbox{$\odot$}}

\begin{document}

\title{Weak lensing evidence for a filament between
  A222/A223\footnote{Based on observations made at ESO/La Silla under
    program Nos. 064.L-0248, 064.O-0248, 66.A-0165, 68.A-0269.}}
\author{J. P. Dietrich\thanks{dietrich@astro.uni-bonn.de},
  D. Clowe, P. Schneider \\
  {\small IAEF, University of Bonn}\\
  J. Kerp\\
  {\small Radioastronomisches Institut, University of Bonn}\\
  E. Romano-D{\'\i}az\\
  {\small Kapteyn Institute, University of Groningen}
}
\date{}
\maketitle
\begin{abstract}
  We present a weak lensing analysis and comparison to optical and
  X-ray maps of the close pair of massive clusters A222/223.
  Indications for a filamentary connection between the clusters are
  found and discussed.
\end{abstract}

\section{Introduction}
\label{sec:introduction}
$N$-body simulations of cosmic structure formation predicts that
matter in the universe should be concentrated along sheets and
filaments and that clusters of galaxies form where these intersect
\citep{1999MNRAS.303..188K,1996Nature..380..603B}. This filamentary
structure, often also dubbed ``cosmic web'', has been seen in galaxy
redshift surveys \citep{1994ApJ...420..525V} and X-rays
\citep{2002A&A...394....7Z,2001ApJ...563..673T}.

Because of the greatly varying mass--to--light $(M/L)$ ratios between
rich clusters and groups of galaxies
\citep{1998Proc..astro-ph/9810298T} it is problematic to convert the
measured galaxy densities to mass densities without further
assumptions. Weak gravitational lensing, which is based on the
measurement of shape and orientation parameters of faint background
galaxies (FBG), is a model--independent method to determine the
surface mass density of clusters and filaments. Due to the finite ellipticities of the unlensed FBG, every weak lensing mass reconstruction
is unfortunately an inherently noisy process, and the expected surface
mass density of a single filament is too low to be detected with
current telescopes \citep{2000ApJ...530..547J}.

Cosmic web theory also predicts that the surface mass density of a
filament increases towards clusters \citep{1996Nature..380..603B}.
Filaments connecting neighboring clusters should have surface mass
densities high enough to be detectable with weak lensing
\citep{1998wfsc.conf...61P}. Such filaments may have been detected in
several recent weak lensing studies.

\citet{1998astro.ph..9268K} found a possible filament between two of
the three clusters in the $z=0.42$ super-cluster MS0302+17, but the
detection remains somewhat uncertain because of a possible foreground
structure overlapping the filament and possible edge effects due to
the gap between two of the camera chips lying along the filament.
Also, \citet{2004astro.ph..1403G} could not confirm the detection of a
filament in this system.  \citet{2002ApJ...568..141G} claim to have
found a filament extending between two of the three clusters of the
Abell~901/902 super-cluster, but the significance of this detection is
low and subject to possible edge effects, as again the filament is
close the intersection of four chips of the camera.

\section{Data of the Abell 222/223 system}
\label{sec:abell-222223-system}
A~222/223 are two Abell clusters at $z
\approx 0.21$ separated by $\sim14\arcmin$ on the sky, or
$\sim2600h_{70}^{-1}$ kpc, belonging to the
\citet{1983ApJS...52..183B} photometric sample. Both clusters are rich,
having Abell richness class 3 \citep{1958ApJS....3..211A}. While these
are optically selected clusters, they have been observed by ROSAT
\citep{1997MNRAS.292..920W,1999ApJ...519..533D} and are confirmed to
be massive clusters. \citet[][P00]{2000A&A...355..443P} published a
list of 53 spectra in the field of A~222/223, 4 of them in region
between the clusters (hereafter ``intercluster region'') and at the
redshift of the clusters. Later, \citet[][D02]{2002A&A...394..395D}
reported on spectroscopy of 183 objects in the cluster field, 153 being
members of the clusters or at the cluster redshift in the intercluster
region. Taking the data of P00 and D02 together, 6 galaxies at the
cluster redshift are known in the intercluster region, establishing
this cluster system as a good candidate for a filamentary connection.

Imaging was performed with the Wide Field Imager (WFI) at the ESO/MPG
2.2 m telescope. In total, twenty 600 s exposures were obtained in
$R$-band in October 2001 centered on A223, eleven 900 s $R$-band
exposures were taken in December 1999 centered on A~222. The images
were taken with a dithering pattern filling the gaps between the chips
in the co-added images of each field.

The $R$-band data used for the weak lensing analysis is supplemented
with three 900 s in the $B$ and $V$-band centered on each cluster
taken from November 1999 to December 2000. The final $B$- and $V$-band
images have some remaining gaps and regions that are covered by only
one exposure and -- due to the dithering pattern -- do not cover
exactly the same region as the $R$-band images.

%
%
%
\subsection{Catalog creation}
\label{sec:catalog-creation}
Lensing catalogs were created from the $R$-band images using the KSB
algorithm \citep[][]{1995ApJ...449..460K}. Stellar reflection rings
and diffraction spikes were masked and the masked regions excluded
from the catalogs. The catalogs from both pointings were merged and
the ellipticities of objects contained in both catalogs were averaged.

From this merged catalog all objects with $R<22$~mag, 
signal-to-noise $\nu<7$, Gaussian radius $0\farcs{4} < r_\mathrm{g} <
1\farcs{2}$, and ellipticity $e > 0.8$ were deleted. All surviving
objects fainter than $R>23.5$ mag were kept as likely background
galaxies, while objects brighter than this -- for which colors were
available -- were selected according to the following criteria:
Objects detected in $B$, $V$, and $R$ matching colors of galaxies at
$z<0.5$, $-0.23 < (V-R) - 0.56 \times (B-V) < 0.67$, $0.5 < B-V <
1.6$, were deleted from the sample. Galaxies detected only in the $V$
and $R$ band were kept if $(V-R) > 1.0$. The final catalog contains
25583 objects, corresponding to a galaxy number density of 15.5
arcmin$^{-2}$ if the masked regions are not taken into account for the
computation of the total area.

\section{Mass and light maps}
\label{sec:weak-lensing-maps}
\begin{figure}[htbp]
  \centering
  \includegraphics[width=\textwidth]{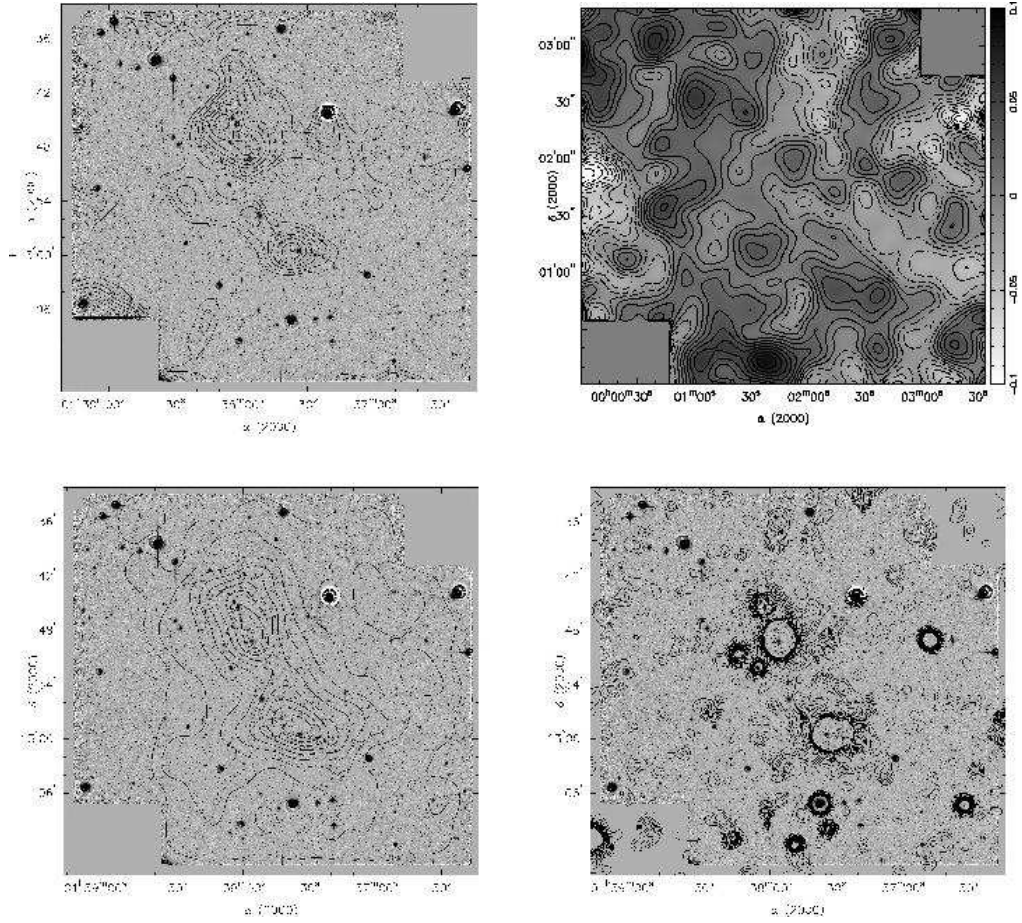}
  \caption{Various light and mass maps of the A~222/223
    system. A~223 is the Northern cluster, A~222 is South of it.
    \emph{Top left}: Weak lensing reconstruction of A222/223.  Each
    contour represents an increase in $\kappa$ of 0.01 ($\sim 3.2
    \times 10^{13}~h_{70}~M_{\sun}$~Mpc$^{-2}$, assuming
    $\overline{z}_\mathrm{FBG} = 1$) above the mean $\kappa$ at the
    edge of the field. The peak $\sim 12\arcmin$ SE of A~222 has an optical
    counter-part.  \emph{Top right}: Surface mass density map created
    from a catalog with randomly rotated galaxies used to estimate the
    noise level of the reconstruction on the right.  \emph{Bottom
      left}: Density distribution of galaxies matching the colors of
    the red cluster sequence ($0.78< V-R < 0.98$) of the Abell
    clusters smoothed with a $2\arcmin$ Gaussian. The absence of
    galaxies in the Eastern part is caused by the very bad or missing
    $V$-band coverage. \emph{Bottom right}:
    Contours in this plot are from ROSAT PSPC data in the energy range
    $0.5 - 2.4$~keV.  The lowest contour is at the $6\sigma$ level.
    Higher contours rise in steps of $1\sigma$. Note that the extended
    X-ray emission North of A~223 is associated with the Northern
    extension of A~223 in the surface mass density map.}
  \label{fig:maps.eps}
\end{figure}

Fig. \ref{fig:maps.eps} shows a reconstruction obtained from the
catalog described in the previous section. The reconstruction was
performed on a $214\times200$ points grid using the algorithm of
\citet{2001A&A...374..740S} adapted to the field geometry. The
smoothing scale of the shear data was set to $2\arcmin$. Galaxies in
the catalog were weighted by the error estimate of their initial
ellipticity measurement.

Both clusters are well detected in the reconstruction. The two
components of A~223 are clearly visible. The strong mass peak West of
A~223 is most likely associated with the reflection ring around the
bright $V=7.98$~mag star at that position. Although the reflection
ring and a large area around it were masked, diffuse stray light is
visible extending beyond the masked region, well into A~223, probably
being the cause of the observed mass peak.

Also visible is a bridge in the surface mass density extending between
A~222 and A~223. Although the signal of this possible filamentary
connection between the clusters is very low, the feature is quite robust when
the selection criteria of the catalog are varied and it never disappears.

The noise level in the reconstruction can be estimated by randomly
rotating the galaxies while keeping their positions and ellipticity
moduli fixed. Performing a reconstruction on this randomized catalog
gives $\left<\kappa^2\right> \simeq 0.015$, suggesting that the
intercluster connection is present at the $\sim 2\sigma$ level.

The galaxy density distribution and the X-ray contours in
Fig. \ref{fig:maps.eps} both show a clear connection between the
clusters. While the X-ray and galaxy density contours are aligned in
the intercluster region, the surface mass density contours connect the
clusters Eastwards of them.

\section{Discussion}
\label{sec:discussion}
We presented weak lensing, optical galaxy density, and X-ray maps of
the massive, close pair of galaxy clusters A~222/223. All maps show a
connection between the clusters in the intercluster region. The
density distribution of color-selected objects, supplemented by the
spectroscopic confirmation of galaxies at the cluster redshift in the
intercluster region by P00 and D02, and the significant connection
between both clusters in the $0.5 - 2.4$~keV band of ROSAT's PSPC
establish a secure filamentary connection between the clusters. The
weak lensing indications for a dark matter bridge are much less
secure. The signal-to-noise of the structure extending between the
clusters is low but the structure itself is very robust and never
disappears when the selection criteria for the lensing catalog are
varied.

We note that the luminosity density of color-selected galaxies in the
intercluster region is only by a factor of $\sim 2$ lower than in the
peaks of the galaxy clusters. If the early type galaxies trace the
mass well and a constant $M/L$ ratio is assumed, we would expect a
clearly detectable lensing signal along the galaxy density contours.
This is obviously not the case. A more detailled comparison of the
observed and expected surface mass density, also addressing the
apparent misalignment of mass and light in A~223, will be done
elsewhere (Dietrich et al. 2004, in preparation). We also note that a
misalignment of mass and light is present in the filament candidate of
\citet{2002ApJ...568..141G}.

\vspace{0.5cm}\noindent 
\emph{Acknowledgment}: This work has been supported by the German
Ministry for Science and Education (BMBF) through DESY under the
project 05AE2PDA/8, and by the Deutsche Forschungsgemeinschaft under
the project SCHN 342/3--1.

\bibliography{a222+3}

\end{document}